\documentclass[aps,pra,superscriptaddress,reprint,floatfix,longbibliography,showpacs]{revtex4-1}
\usepackage[english]{babel}
\usepackage{amsmath}
\usepackage{amssymb}
\usepackage{graphicx}
\usepackage{hyperref}
\usepackage{upgreek}
\usepackage{epstopdf}
\usepackage{color}
\usepackage{dcolumn}
\usepackage{multirow}
\usepackage[note-name=, use-sort-key = false]{notes2bib}

\def\cm{cm$^{-1}$}

\begin{document}
\title{Infrared study of the layered, magnetic insulator Mn(Bi$_{0.07}$Sb$_{0.93}$)$_2$Te$_4$ at low temperatures}
\author{M. K\"opf}
\affiliation{Experimentalphysik II, Institute of Physics, Augsburg University, 86159 Augsburg, Germany}
\author{S. H. Lee}
\affiliation{2D Crystal Consortium, Materials Research Institute, Pennsylvania State University, University Park, PA 16802, USA}
\affiliation{Department of Physics, Pennsylvania State University, University Park, Pennsylvania 16802, USA}
\author{H. Kumar}
\affiliation{Experimentalphysik II, Institute of Physics, Augsburg University, 86159 Augsburg, Germany}
\author{Z. Q. Mao}
\affiliation{2D Crystal Consortium, Materials Research Institute, Pennsylvania State University, University Park, PA 16802, USA}
\affiliation{Department of Physics, Pennsylvania State University, University Park, Pennsylvania 16802, USA}
\affiliation{Department of Materials Science and Engineering, Pennsylvania State University, University Park, Pennsylvania 16802, USA}
\author{C. A. Kuntscher}\email{christine.kuntscher@physik.uni-augsburg.de}
\affiliation{Experimentalphysik II, Institute of Physics, Augsburg University, 86159 Augsburg, Germany}

\begin{abstract}
Topological insulators with intrinsic magnetic ordering, potentially hosting rare quantum effects, recently attracted extensive attention. MnBi$_2$Te$_4$ is the first established example. The Sb-doped variant Mn(Bi$_{1-x}$Sb$_x$)$_2$Te$_4$ shows a great variety of electronic properties depending on the Sb content~$x$, such as shifts in the Fermi level and the Neel temperature $T_{\mathrm N}$, and the change of the free charge carrier type from $n$- to $p$-type at high Sb substitution ratios.
Here, we investigate the effect of magnetic ordering on the bulk electronic structure of Mn(Bi$_{1-x}$Sb$_x$)$_2$Te$_4$ with high Sb content $x=0.93$ by temperature-dependent reflectivity measurements over a broad frequency range. We observe anomalies in the optical response across $T_{\mathrm N}$ when the antiferromagnetic order sets, which suggests a coupling between the magnetic ordering and the electronic structure of the material.
\end{abstract}

\maketitle

\section{Introduction}
\begin{figure}[t]
	\includegraphics[width=1\linewidth]{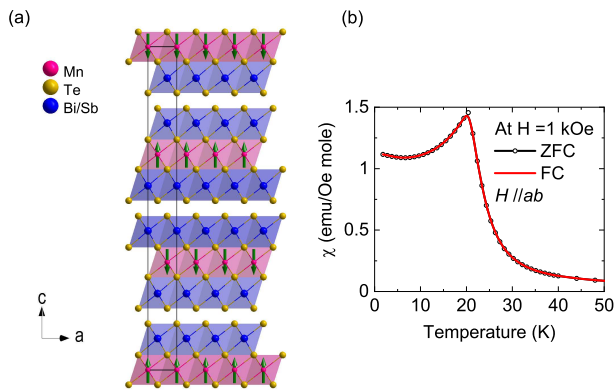}
	\caption{\label{fig.Structure} (a) Sketch of the crystal structure and unit cell of Mn(Bi$_{1-x}$Sb$_x$)$_2$Te$_4$ in the $A$-type antiferromagnetic state. The arrows indicate the orientation of the Mn spins. (b) Magnetic susceptibility of Mn(Bi$_{0.07}$Sb$_{0.93}$)$_2$Te$_4$ with an anomaly at 20\,K indicating the antiferromagnetic ordering transition.}
\end{figure}

The topological insulator MnBi$_2$Te$_4$ (MBT) is currently extensively investigated, since its intrinsic antiferromagnetic order could potentially induce interesting quantum mechanical effects such as quantum anomalous Hall effect (QAH) or axion insulator at low temperature~\cite{Li.2020, Fang.2021}.
The QAH, in particular, plays a crucial role for potential applications in quantum metrology and spintronics~\cite{Lei.2021}. In recent studies, reduced dimensionality proved to realize exotic phenomena in van-der-Waals-type layered materials, which favours the research of thin films/monolayers of this material family~\cite{An.2021}.
MBT belongs to the group of ternary chalcogenides and its layered crystal structure with space group $R\bar{3}m$ is built of septuple layers, where MnTe blocks are intercalated in Bi$_2$Te$_3$ layers.
By exchanging certain elements gradually -- in this case bismute (Bi) by antimone (Sb) -- the electronic and magnetic properties can be manipulated~\cite{Chen.2019, Wang.2021}. Furthermore, an ideal type-II Weyl semimetal can be established through appropriate Sb doping in high magnetic fields \cite{Lee.2021}. The unit cell parameters are slightly changing in Mn(Bi$_{1-x}$Sb$_x$)$_2$Te$_4$  from $a = 4.33\,$\AA\  and $c = 40.93\,$\AA\ for $x=0$ to $a = 4.25\,$\AA\  and $c = 40.87\,$\AA\ for $x=1$ \cite{Yan.2019a}.
Fig.\,\ref{fig.Structure}\,(a) displays the unit cell structure of the mixed compound Mn(Bi$_{1-x}$Sb$_x$)$_2$Te$_4$, where the septuple layer interaction is of van-der-Waals type~\cite{Li.2020a}.
In the antiferromagnetic phase, the out-of-plane Mn spins are aligned parallel within the $ab$ plane and anti-parallel along the $c$ axis. This $A$-type antiferromagnetic state in Mn(Bi$_{1-x}$Sb$_x$)$_2$Te$_4$ orders between $T_{\mathrm N} = 25\,$\,K ($x=0$) and $T_{\mathrm N} = 19\,$\,K ($x=1$)~\cite{Yan.2019a}. In case of a high Sb content, also ferrimagnetism has been observed, which possibly results from anti-site mixing of Mn and Sb ions~\cite{Riberolles.2021}.

According to magnetic susceptibility measurements, the effective magnetic moment $\upmu_{\mathrm {eff}} = 5.3\,\upmu_{\mathrm B}$, originating from the Mn$^{2+}$ ions, is not changed by the Sb doping ratio~\cite{Yan.2019a}.
However, the magnetic properties of the compounds are significantly affected, namely, the antiferromagnetic ordering temperature, the saturation moment, Weiss constant, and critical fields $H_{\mathrm c}$ for the spin flop transition all decrease with increasing Sb content \cite{Yan.2019a,Lee.2021}. Additionally, a strong influence of Sb doping on the electronic structure can be found, as the free charge carrier type is changing from $n$-type to $p$-type at a ``critical'' doping level of $x\approx0.26$ \cite{Chen.2019,Yan.2019a,Lee.2021}.
It was furthermore predicted that at $x=0.55$ the energy gap closes and reopens, and that a topological phase transition occurs, where the material turns from an intrinsic magnetic topological insulator with an inverted band gap to a topologically trivial magnetic insulator without band inversion \cite{Chen.2019,Ko.2020}.
This scenario was, however, contradicted by a recent theoretical work which showed that the energy gap is reduced from 138~meV (for MnBi$_2$Te$_4$) to 16~meV (for MnSb$_2$Te$_4$) due to the reduced spin-orbit coupling, but remains inverted~\cite{Li.2021}. According to Ref.~\cite{Li.2021} both MnBi$_2$Te$_4$ and MnSb$_2$Te$_4$ are therefore expected to be topological insulators.

\begin{figure}[t]
	\includegraphics[width=\linewidth]{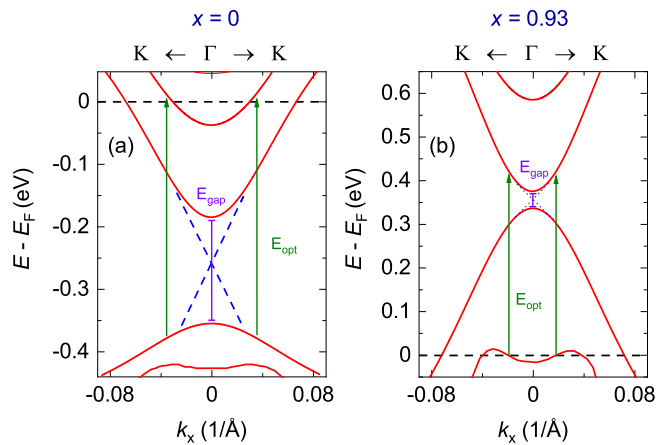}
	\caption{\label{fig.BandStructure} Sketch of the electronic structures of (a) MnBi$_2$Te$_4$ and (b) Mn(Bi$_{0.07}$Sb$_{0.93}$)$_2$Te$_4$ near the band gap and Fermi level based on Refs.\ \cite{Chen.2019,Chen.2019a,Li.2021}. The blue dashed and dotted lines indicate the (topological) surface states in these compounds. The optical gap and the energy gap are shown with green and violet lines, respectively.}
\end{figure}

Despite the disagreement regarding the topological character of the Mn(Bi$_{1-x}$Sb$_x$)$_2$Te$_4$ materials, the reports agree in that the energy gap is decreasing with increasing $x$, and that for a high Sb content $x>0.55$, a metallic behavior with mainly $p$-type charge carriers prevails.
A sketch of the bulk band structure for MnBi$_2$Te$_4$ and Mn(Bi$_{0.07}$Sb$_{0.93}$)$_2$Te$_4$ (MBST) studied in this work is shown in Fig.\,\ref{fig.BandStructure}. It reflects the evolution of the electronic bands near the Fermi level and the (topological) surface states with increasing Sb content. Since the topological properties of highly doped Mn(Bi$_{1-x}$Sb$_x$)$_2$Te$_4$ is controversial, the surface states are indicated by dotted lines in Fig.\,\ref{fig.BandStructure}\,(b). Regarding the bulk band structure of the measured compound MBST, the Fermi level is pushed down and crosses the former valence bands resulting in mainly $p$-type free charge carriers, and the band gap is reduced compared to the pure compound~\cite{Chen.2019}.
Due to the Fermi level crossing of the electronic bands,
the MBST sample is expected to show signs of a metallic character with a high reflectivity at low energies, similar to the results for the pure compound \cite{Koepf.2020,Xu.2021}.

In this work, we investigate the optical excitations in Mn(Bi$_{0.07}$Sb$_{0.93}$)$_2$Te$_4$ by temperature-dependent reflectivity measurements over a broad frequency range, in order to characterize the changes in the electronic structure induced by the magnetic phase transition. The obtained results are compared to the recent reports \cite{Koepf.2020,Xu.2021} on the undoped material MBT.

\section{Methods}

Single crystals of Mn(Bi$_{0.07}$Sb$_{0.93}$)$_2$Te$_4$ were grown by the self-flux method as reported in Ref.\ \cite{Yan.2019}. The plate-like sample had a surface size of app. 0.6 x 0.8\,mm and a thickness close to 100\,$\upmu$m. Magnetic susceptibility data have been collected from 1.8\,K to 300\,K using a superconducting quantum interference device (SQUID, Quantum Design) magnetometer MPMS. According to these measurements, the sample undergoes a magnetic phase transition at $\sim20\,$K, as expected for $x=0.93$[see Fig.\,\ref{fig.Structure}\,(b)]. The magnetic field $H$ has been aligned parallel of the $ab$-plane, i.e., perpendicular to the antiferromagnetic ordering.
For the reflectivity measurements at temperatures between 295 and 5\,K, we have used a CryoVac Konti cryostat, which has been connected to a Bruker Hyperion infrared microscope and Bruker Vertex80v FTIR spectrometer. Half of the surface of the freshly cleaved sample was coated with a thin silver layer, which was used as reference for the calculation of the absolute reflectivity. The sample was glued to a sample holder within the cryostat and aligned perpendicular to the incoming beam. The measurements were performed from the far-infrared up to the visible range (100 to 20000\,\cm). The measured spectra were extrapolated in the low- and high-frequency range with the help of literature values and volumetric data. Then, the optical functions were calculated through the Kramers-Kronig relations, using programs by David Tanner~\cite{Tanner.2015}. The optical spectra were fitted with the Drude-Lorentz model using the software RefFIT~\cite{Kuzmenko.2005}.

\section{Results and Discussion}

\begin{figure*}[t]
	\includegraphics[width=\linewidth]{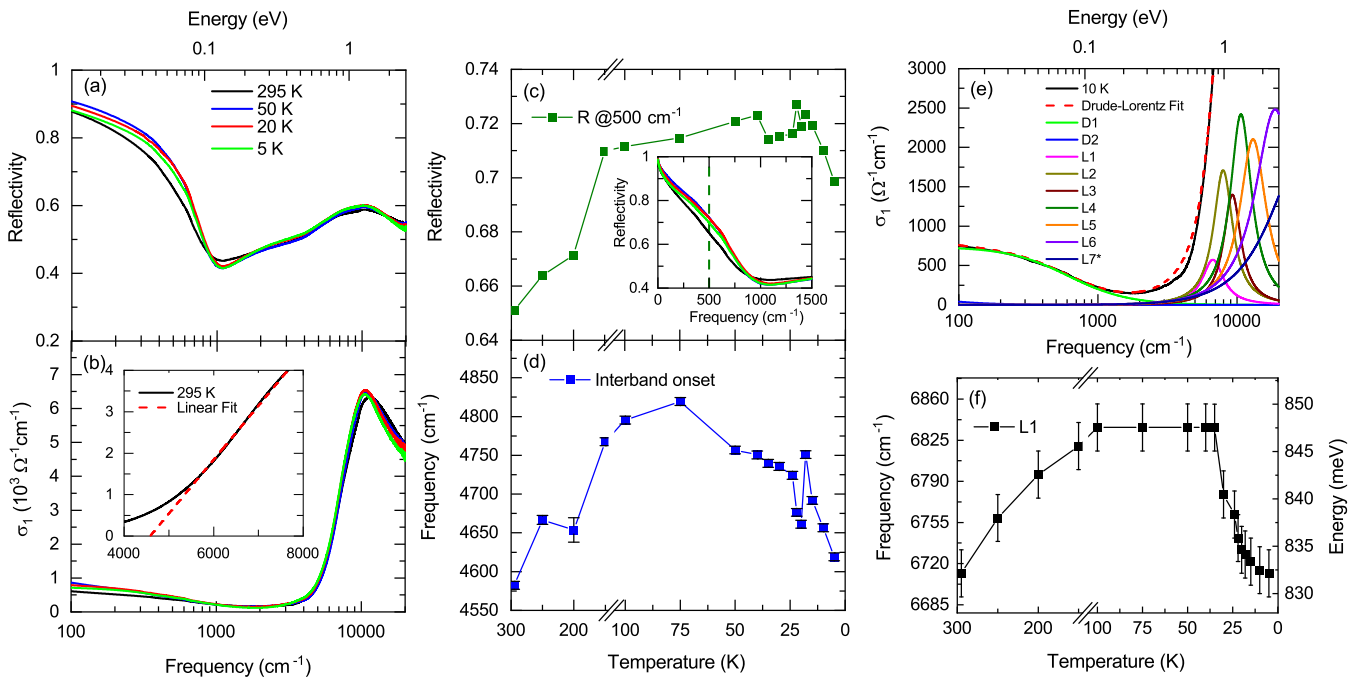}
	\caption{\label{fig.optics}
(a) Reflectivity and (b) optical conductivity $\sigma_1$ of Mn(Bi$_{0.07}$Sb$_{0.93}$)$_2$Te$_4$ for selected temperatures. (c) Reflectivity values at wavenumber 500\,\cm (inset shows the relevant range) for the measured temperatures. (d) Interband onset values at low temperatures resulting from the zero-crossing of the linear extrapolations in $\sigma_1$ [see inset in (b)]. (e) Drude-Lorentz fit of the optical conductivity at 10\,K with two Drude (D) and seven Lorentz (L) oscillators. (f) Temperature dependence of the L1 oscillator position.
}
\end{figure*}

The reflectivity spectra of MBST are shown in Fig.\,\ref{fig.optics}\,(a) for selected temperatures. The temperature steps have been decreased close to the phase transition temperature $T_{\mathrm{N}}\approx 20\,$K. The high reflectivity at low frequencies and the plasma edge near 1000\,\cm\ indicate the metallic character of the material. The bumps in the reflectivity spectra above $\sim$1000~cm$^{-1}$ are due to electronic transitions across the optical gap.
With decreasing temperature, slight but significant changes can be observed in the low-energy range: During cooling from 295 to 50\,K the plasma edge sharpens and the Drude spectral weight, which is associated with the free charge carriers, increases. This trend is, however, reversed for temperatures below 50~K. In the high-frequency range, the temperature-induced effects in the reflectivity spectrum appear to be weak, however, as we will see below, are significant.
For further illustration of the temperature-induced changes, we plot in Fig.\,\ref{fig.optics}\,(c) the reflectivity values at 500\,\cm, i.e., within the free charge carrier range, as a function of temperature. During cooling, one observes a steady increase of this value down to 40\,K, followed by slight deviations down to 20\,K and a decrease of the reflectivity at very low temperatures, which might already be a hint for the magnetic ordering.

The optical conductivity spectrum $\sigma_1$, which was obtained from the measured reflectivity via KK analysis, is depicted in Fig.\,\ref{fig.optics}\,(b). The excitations of the free charge carriers are visible from 0 up to about 2000\,\cm, where we detected the plasma minimum, separating the intraband from the interband transition range. The latter starts from about 4000\,\cm, where a steep linear increase in $\sigma_1$ is visible. This region is followed by a broad maximum at $\sim$12000\,\cm. Thus, the Drude spectral weight is rather small compared to the spectral weight of the optical transitions at higher energies, which indicates a relatively weak metallic behavior with $\sigma_{\mathrm dc}$ values smaller than 1000\,$\Omega^{-1}$\cm. The shape of $\sigma_1$ is comparable to the one we measured for the undoped compound MBT, where the spectral weight of free charge carriers is slightly higher, yet the plasma minimum and the interband transition onset are located at lower energies~\cite{Koepf.2020}.

With the use of linear approximation of the optical conductivity
$\sigma_1$ [see the example in the inset of Fig.\,\ref{fig.optics}\,(b)] we roughly
estimate the onset of the interband transitions at each temperature
and associate it with the optical gap $E_{\mathrm{opt}}$.
The energy $E_{\mathrm{opt}}$ corresponds to electronic transitions from the highest occupied states below the Fermi level to the next higher lying band across the energy gap $E_{\mathrm{gap}}$, while $E_{gap}$ corresponds to the smallest energy difference between the valence and conduction band [see Fig.\,\ref{fig.BandStructure}\,(b)]. The difference between $E_{\mathrm{opt}}$ and $E_{\mathrm{gap}}$ is due to the Moss-Burstein shift~\cite{Burstein.1954}, which describes the (de)population of states in the conduction (valence) band and the subsequent shift of the Fermi level. The so-obtained values of $E_{\mathrm{opt}}$ are plotted in Fig.\,\ref{fig.optics}\,(d): during cooling, an increase of the optical gap occurs down to  75\,K, followed by a moderate decrease down to 20\,K.
At 18\,K one observes a sudden jump to higher values, followed by a further steady decrease. This jump might be caused by the magnetic phase transition occurring at this temperature in our sample, since a blue shift of the energy gap can be induced by the onset of magnetic ordering as it has been reported for MnTe~\cite{Bossini.2020}. We assume, that also the optical gap can be affected by this blue shift, which should be visible in our results.

\begin{figure*}[t]
	\includegraphics[width=\linewidth]{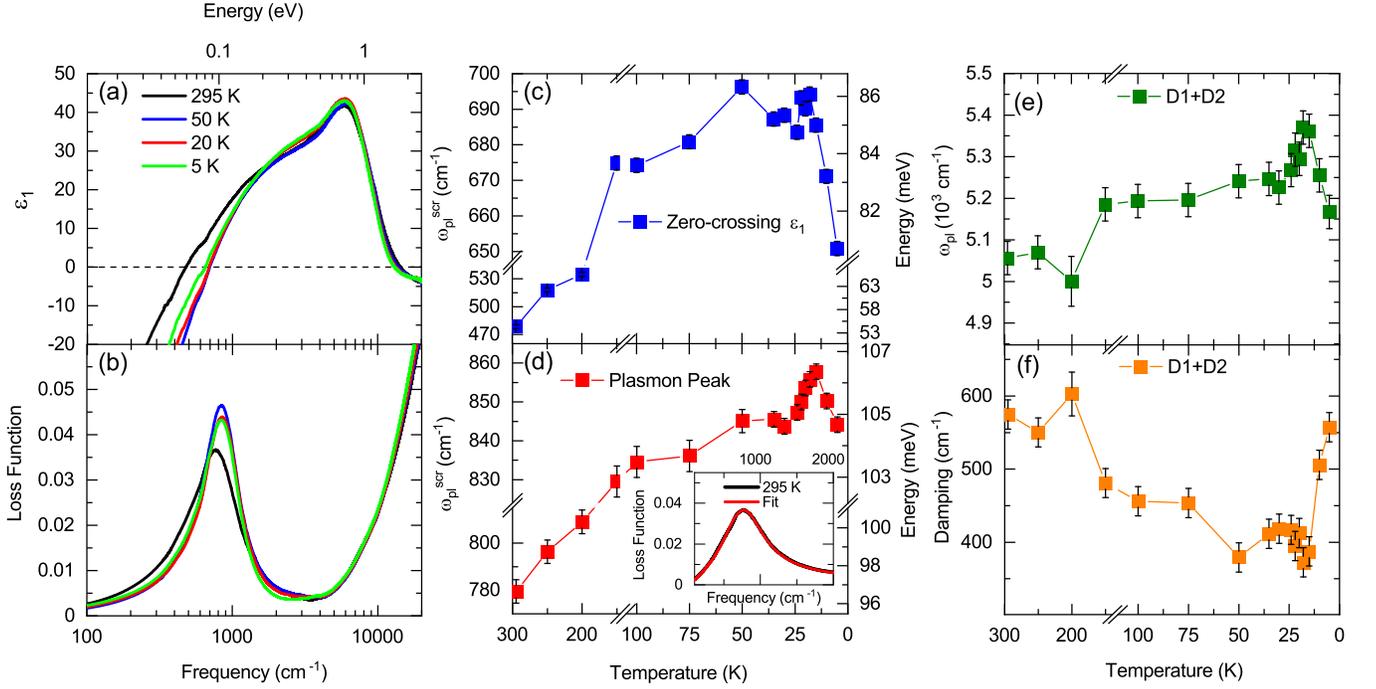}
	\caption{\label{fig.D1+D2} (a) $\varepsilon_1$ and (b) loss function of Mn(Bi$_{0.07}$Sb$_{0.93}$)$_2$Te$_4$ at selected temperatures. (c) and (d) show the temperature-dependent values of the screened plasma frequency $\omega^{\mathrm{scr}}_{\mathrm{pl}}$ resulting from the zero-crossing of $\varepsilon_1$ and the peak position of the loss function, respectively. (e) and (f) display the temperature dependence of the plasma frequency $\omega_{\mathrm{pl}}$ and damping of the combined two Drude terms from the Drude-Lorentz fitting.}
\end{figure*}

For a quantitative analysis, we performed a simultaneous fitting of the reflectivity and optical conductivity spectra with the Drude-Lorentz model [see Fig.\,\ref{fig.optics}\,(e)], where we used the same number of oscillators like for the undoped compound MnBi$_2$Te$_4$ \cite{Koepf.2020}.
Two Drude terms have been implemented to characterize the response of the free charge carriers, which should be mainly $p$-type according to Chen et al.~\cite{Chen.2019}. Concerning the spectral weight, we have used one strong and one weak Drude term, as shown in Fig.\,\ref{fig.optics}\,(e).
This can be justified, like for the undoped compound \cite{Koepf.2020,Xu.2021}, by the free carrier contributions of two different conduction bands. In Fig.\,\ref{fig.BandStructure} the band structure of MBST is sketched and the crossing of the Fermi level with two different bands is demonstrated. Due to the higher density of states of one band compared to the other, one Drude term has a much larger spectral weight than the other. The temperature-dependent values of the position of the L1 oscillator are summarized in Fig.\,\ref{fig.optics}\,(f). This term is located near the onset of the interband transitions and, therefore, it can be associated with the temperature-dependent evolution of the optical gap.
Consistently, we find similarities between the temperature dependence of the L1 frequency and that of the interband onset depicted in Fig.\,\ref{fig.optics}\,(d), namely an anomalous behavior close to the magnetic phase transition temperature.

Besides the optical conductivity, also other optical functions show an anomaly in their temperature dependence.
The real part of the dielectric function $\varepsilon_1$ and the loss function, which is defined
as Im(-1/$\varepsilon(\omega)) = \varepsilon_2/(\varepsilon_1^2+\varepsilon_2^2)$ where $\varepsilon(\omega)$ is the complex dielectric function, plotted in Figs.\ \ref{fig.D1+D2}\,(a) and (b), respectively, provide insight into the temperature dependence of the screened plasma frequency $\omega^{\mathrm{scr}}_{\mathrm{pl}}$.
Regarding the $\varepsilon_1$ function, the value of $\omega^{\mathrm{scr}}_{\mathrm{pl}}$ is given by the frequency of the zero-crossing, while from the loss function this value can be extracted from the position of the plasmon peak. In Fig.\,\ref{fig.D1+D2}\,(c) and (d), these values are plotted as a function of temperature. The values derived from $\varepsilon_1$ show an increase between 295 and 50\,K from $\sim$480\,\cm\ to almost 700\,\cm. After a mainly constant behaviour down to 20\,K, the values are dropping to  $\sim$650\,\cm\ at 5\,K. A similar trend is seen for the temperature dependence of $\omega^{\mathrm{scr}}_{\mathrm{pl}}$ determined from the plasmon peak Lorentz fit [see inset of Fig.\,\ref{fig.D1+D2}\,(d)]. With decreasing temperature the values rise steadily down to 30\,K from  $\sim$780 to 845\,\cm, whereafter a much stronger increase follows up to 860\,\cm\ at 18\,K. With further cooling, $\omega^{\mathrm{scr}}_{\mathrm{pl}}$ is shifting to lower frequencies to 843\,\cm\ at 5\,K, similar to the trend seen in Fig.\,\ref{fig.D1+D2}\,(c). In general, both ways of specifying $\omega^{\mathrm{scr}}_{\mathrm{pl}}$ should agree with each other, but the absolute values can differ by a certain value, which might be caused by the effect of high-energy transitions, as described in~\cite{Wooten.1972}. Yet, we can find a very good agreement regarding the temperature dependence and the cusp near the phase transition temperature $T_{\mathrm N}=20\,$K.

In Fig.\,\ref{fig.D1+D2}\,(e) and (f) we plot the plasma frequency $\omega_{\mathrm{pl}}$ and the damping $D$ of the two Drude terms, resp.\ , as extracted from the fitting [see Fig.\,\ref{fig.optics}\,(e)] and calculated according to the equations \cite{Dressel.2002}
\begin{equation}
    \omega_{pl} = \sqrt{\omega_{pl,1}^2+\omega_{pl,2}^2}
\end{equation}
\begin{equation}
D = \frac{\sqrt{\omega_{pl,1}^2+\omega_{pl,2}}^2}{8\pi^2c\sigma_{dc}x}  \quad  .
\end{equation}

The plasma frequency can also be expressed by the charge density $N$ and the effective mass $m^*$ with the formula $\omega_{pl} = \sqrt{4\pi N e^2/m^*}$~\cite{Dressel.2002}. $\omega_{pl}$ is increasing steadily from 295\,K down to 18\,K, whereas below 18\,K it decreases, which symbolizes a weakening of the metallic character. This could either originate from a change in the charge carrier density or in the effective mass due to band profile modifications [see Fig.\,\ref{fig.D1+D2}\,(e)].
The damping is decreasing from 295\,K to 18\,K, and below 18\,K we find a strong increase.
Accordingly, the lowering of temperature causes a growth in the metallic characteristics of the sample, but below $T_{\mathrm N}$ this trend is reversed.
Thus, also the parameters $\omega_{\mathrm{pl}}$ and $D$ show an anomaly at approx.\ 20\,K, which agree with the values from Figs.\,\ref{fig.D1+D2}\,(c) and (d).
We also point out that the temperature behaviors of $\omega^{\mathrm{scr}}_{\mathrm{pl}}$ and $\omega_{\mathrm{pl}}$ are in good agreement with each other. Yet, the absolute values differ by a factor, since $\omega^{\mathrm{scr}}_{\mathrm{pl}}$ is affected by high-energy excitations and, hence, shifted to lower values, while $\omega_{\mathrm{pl}}$ is solely determined by the Drude contributions. Quantitatively, the relation
$\omega^{\mathrm{scr}}_{\mathrm{pl}}=\omega_{\mathrm{pl}}/\sqrt{\varepsilon_\infty}$ holds \cite{Wooten.1972}. At the lowest temperature (5\,K) we obtain
$\omega^{\mathrm{scr}}_{\mathrm{pl}}=844$~cm$^{-1}$ (from the loss function) and $\omega_{\mathrm{pl}}=5167$~cm$^{-1}$, which gives the value $\varepsilon_\infty=37.5$. We extract a similar value ($\sim$43) from the function $\varepsilon_1$ close to its maximum at $\sim$6000\,\cm. Hence, $\varepsilon_\infty$ characterizes the interband transitions at higher energies.

The observed weakening of the free charge carrier density at $T_\mathrm{N}$ could be related to the opening of a gap in the electronic surface states when the antiferromagnetic order sets in. He et al.~\cite{He.2020} discussed an opening of the topological surface gap in the case of pure MBT when the magnetic moments start to align in an $A$-type antiferromagnetic order. Although infrared spectroscopy is mainly bulk and not surface sensitive, a contribution of surface states, especially during the onset of magnetic order, has been reported in recent studies~\cite{Xu.2021a, Regmi.2020}.

As discussed by Bossini {\it et al.}~\cite{Bossini.2020}, Xu {\it et al.}~\cite{Xu.2021a}, and Dalui {\it et al.}~\cite{Dalui.2021}
regarding the materials MnTe, EuIn$_2$As$_2$, and (Sb$_{0.95}$Cr$_{0.05}$)$_2$Te$_3$, respectively,
the magnetic phase transition affects the bulk electronic structure near the Fermi level, and thus the intra- or interband excitations at low energies.
Among these studies, an interaction of the AFM spin fluctuations with the effective mass has been observed, which caused a cusp-like anomaly in the plasma frequency $\omega_{\mathrm{p}}$ at the phase transition temperature $T_{\mathrm{N}}$~\cite{Xu.2021a}. Also, the plasma frequency from the surface states is expected to be very temperature dependent and to be affected by the AFM transition~\cite{Xu.2021a}. Besides the breaking of the time-reversal symmetry for the realization of exotic quantum effects, the creation of an exchange gap at the Dirac point also turned out to be a consequence of the presence of ordered magnetic moments~\cite{Dalui.2021, Tokura.2019}, which might be visible in our results. Even though Mn(Bi$_{1-x}$Sb$_x$)$_2$Te$_4$ is expected to be a topologically trivial magnetic insulator for $x$ $>$ 0.55, as described above, recent DFT calculations~\cite{Eremeev.2021} and surface-sensitive ARPES studies, of Mn rich, epitaxal films, revealed topological characteristics for the compound MnSb$_2$Te$_4$~\cite{Wimmer.2021}. This could also be the case for MBST studied here.
In the case of pure MBT, He et al.~\cite{He.2020} discussed the effect of the magnetic ordering on the electronic bands.
A hybridization of the Bi and Te $p$-bands close to the Fermi level, which are responsible for the topological character in the MBST family, and the Mn $d$-bands, which are further away from the Fermi level, could be detected. The antiferromagnetic order affects this hybridization, since below the ordering temperature the ferromagnetic layers are interacting differently compared to the layers above $T_{\mathrm{N}}$. Depending on an even or odd number of magnetic topological layers, a QAH insulator or axion insulator can be created~\cite{He.2020}. As a result, a magnetic gap at the topological surface states (TSS) is generated, which could be a reason for the observed decrease of the free charge carrier density. Since infrared spectroscopy is mainly sensitive to bulk electronic properties and less to surface states, this contribution should be minor, which applies to our findings. Eventually, this might explain the weakening of the metallic character of Mn(Bi$_{0.07}$Sb$_{0.93}$)$_2$Te$_4$ below $T_N$, as revealed by the temperature dependence of $\omega^{\mathrm{scr}}_{\mathrm{pl}}$ and    $\omega_{\mathrm{pl}}$.

\section{Conclusion}
In conclusion, we studied the temperature-dependent optical functions of the magnetic insulator Mn(Bi$_{0.07}$Sb$_{0.93}$)$_2$Te$_4$ by reflectivity measurements, in order to characterize the effect of the antiferromagnetic ordering on the electronic structure near the Fermi level. Similar to the
topological insulator MnBi$_2$Te$_4$ \cite{Koepf.2020,Xu.2021}, we have detected an anomaly in the profile of the spectra and several optical parameters at $T_{\mathrm N} = 20\,$K. From our findings, we conclude an interplay between the magnetic ordering and the electronic structure in Mn(Bi$_{0.07}$Sb$_{0.93}$)$_2$Te$_4$. The anomalous behavior might be caused by the opening of an exchange gap at the surface Dirac point, due to the breaking of the time reversal symmetry when the antiferromagnetic ordering sets in. \\

\begin{acknowledgments}
C.\ A.\ K.\ acknowledges financial support from the Deutsche
Forschungsgemeinschaft (DFG), Germany, through Grant
No. KU 1432/15-1.  Z.Q.M. and S.H.L. acknowledges the support of the US NSF through the Penn State 2D Crystal Consortium-Materials Innovation Platform (2DCC-MIP) under NSF Cooperative Agreement DMR-2039351.
\end{acknowledgments}

\end{document}